\documentclass[pra,showpacs]{revtex4}
\usepackage{amssymb}
\usepackage{amsmath,amssymb}

\begin{document}

\title{Conditional linear-optical measurement schemes generate effective photon nonlinearities}
\author{G.G.\ Lapaire$^1$, Pieter Kok$^2$, Jonathan P.\ Dowling$^2$, and J.E.\ Sipe$^1$}
\affiliation{$^1$Department of Physics, University of Toronto,
60 St.~George St., Toronto, ON M5S 1A7, Canada \\
$^2$Quantum Computing Technologies Group, Section 367, Jet Propulsion Laboratory,
California Institute of Technology, Mail Stop 126-347, 4800 Oak Grove Drive, Pasadena, CA 91109}


\begin{abstract}
 We provide a general approach for the analysis of optical state
 evolution under conditional measurement schemes, and identify the
 necessary and sufficient conditions for such schemes to simulate
 unitary evolution on the freely propagating modes. If such unitary
 evolution holds, an effective photon nonlinearity can be
 identified. Our analysis extends to conditional measurement schemes
 more general than those based solely on linear optics.
\end{abstract}
\maketitle
\section{\protect\smallskip Introduction}

One of the main problems that optical quantum computing has to overcome is
the efficient construction of two-photon gates \cite{kok00}. We can use Kerr
nonlinearities to induce a phase shift in one mode that depends on the
photon number in the other mode, and this nonlinearity is sufficient to
generate a universal set of gates \cite{kerr}. However, passive Kerr media
have typically small nonlinearities (of the order of $10^{-16}\,\text{cm}%
^{2}\,\text{sV}^{-1}$ \cite{boyd99}). We can also construct large Kerr
nonlinearities using slow light, but these techniques are experimentally
difficult \cite{lukin00}.

On the other hand, we can employ linear optics with projective measurements.
The benefit is that linear optical schemes are experimentally much easier to
implement than Kerr-media approaches, but the downside is that the
measurement-induced nonlinearities are less versatile and the success rate
can be quite low (especially when inefficient detectors are involved).
However, Knill, Laflamme and Milburn \cite{klm} showed that with sufficient
ancilla systems, these linear-optical quantum computing (LOQC) devices can
be made near-deterministic with only polynomial resources. This makes linear
optics a viable candidate for quantum computing. Indeed, many linear optical
schemes and approaches have been proposed since \cite
{gottesman,franson,simple,mathis,kyi,snmk}, and significant experimental
progress has already been made \cite{feedfwd,franson2}.

The general working of a device that implements linear optical processing
with projective measurements is shown in Fig.~1. The computational input and
the ancilla systems add up to $N$ optical modes that are subjected to a
unitary transformation $U$, which is implemented with beam splitters, phase
shifters, \textit{etc}. This is called an optical $N$-port device. In order
to induce a transformation of interest on the computational input, the
output is conditioned on a particular measurement outcome of the ancilla
system. For example, one can build a single-photon quantum nondemolition
detector with an optical $N$-port device \cite{kok02}. In general, $N$-port
devices have been studied in a variety of applications \cite{nport}.

The class of such devices of interest here is that in which a unitary
evolution on the computational input is effected. To date these devices have
been proposed and studied on a more-or-less case by case basis. Our approach
is to address this class in a more general way, and identify the conditions
that such a device must satisfy to implement a unitary evolution on the
computational input. Once that unitary evolution is established, an
effective photon nonlinearity associated with the device can be identified.

In this paper, we present necessary and sufficient conditions for the
unitarity of the optical transformation of the computational input, and we
derive the effective nonlinearities that are associated with some of the
more common optical gates in LOQC. We begin section \ref{sec:formalism} by
introducing the formalism. In sections \ref{sec:consec}-\ref{sec:nscon}, we
examine the transformation equation under the assumption that it is unitary.
We show that there are two necessary and sufficient conditions for the
transformation to be unitary and we provide a simple test condition. In
section \ref{sec:post}, we expand the formalism and conditions to include
measurement dependent output processing (see Fig.~2), which is used in
several schemes. In section \ref{sec:examples}, we show how the formalism
can be applied to quantum computing gates. We choose as examples two quantum
gates already proposed, the conditional sign flip of Knill, Laflamme, and
Milburn \cite{klm}, and the polarization-encoded CNOT of Pittman \textit{et
al}.\ \cite{franson}. Our concluding remarks are presented in section \ref
{sec:conclusions}, where we note that our main results extend to devices
where the unitary transformation $U$ is more general than those
implementable with linear optics alone.

\section{The general formalism}

\label{sec:formalism}

We consider a class of optical devices that map the computational input
state onto an output state, conditioned on a particular measurement outcome
of an ancilla state (see Fig.~1). We introduce a factorization of the entire
Hilbert space into a space $\mathcal{H}_{C}$ involving the input computing
channels (\textit{i.e.}, both ``target'' and ``control'' in a typical
quantum gate), and a Hilbert space $\mathcal{H}_{A}$ involving the input
ancilla channels, 
\[
\mathcal{H}=\mathcal{H}_{C}\otimes \mathcal{H}_{A}\; . 
\]
We assume that the input computing and ancilla channels are uncorrelated and
unentangled, so we can write the full initial density operator as $\rho
\otimes \sigma $, where $\rho $ is the initial density operator for the
computing channels, and $\sigma $ the initial density operator for the
ancilla channels.

Let $U$ be the unitary operator describing the pre-measurement evolution of
the optical multi-port device. At the end of this process we have a full
density operator given by $U\left( \rho \otimes \sigma \right) U^{\dagger }$%
. In anticipation of the projective measurement, it is useful to introduce a
new factorization of the full Hilbert space into an output computing space $%
\mathcal{H}_{\bar{C}}$ and a new ancilla space $\mathcal{H}_{\bar{A}}$, 
\[
\mathcal{H=H}_{\bar{C}}\otimes \mathcal{H}_{\bar{A}}\; . 
\]

The Von Neumann projective measurements of interest are described by
projector-valued measures (or PVMs) of the type $\left\{ \bar{P},{I-}\bar{P}%
\right\} $, where ${I}$ is the identity operator for the whole Hilbert
space, and the projector $\bar{P}$ is of the form 
\begin{equation}
\bar{P}={I}_{\bar{C}}\otimes \sum_{\bar{k}}s_{\bar{k}}\left| \bar{k}%
\right\rangle \left\langle \bar{k}\right| \; ,  \label{projector}
\end{equation}
where ${I}_{\bar{C}}$ is the identity operator in $\mathcal{H}_{\bar{C}}$,
and we use Roman letters with an overbar, \textit{e.g., }$\left| \bar{k}%
\right\rangle $, to label a set of orthonormal states, $\left\langle \bar{k}|%
\bar{l}\right\rangle =\delta _{\bar{k}\bar{l}}$, spanning the Hilbert space $%
\mathcal{H}_{\bar{A}}$; each $s_{\bar{k}}$ is equal to zero or unity. The
number of nonzero $s_{\bar{k}}$ identifies the rank of the projector $\bar{P}
$ in $\mathcal{H}_{\bar{A}}$. ``Success'' is defined as a measurement
outcome associated with the projector $\bar{P}$, and the probability of
success is thus 
\begin{equation}
d(\rho )\equiv \mathrm{Tr}_{\bar{C},\bar{A}}\left( U\left( \rho \otimes
\sigma \right) U^{\dagger }\bar{P}\right)\; .  \label{Ddef}
\end{equation}
Clearly, in general $d(\rho )$ depends on the ancilla density operator $%
\sigma $, the unitary evolution $U$, and the projector $\bar{P}$, as well as
on $\rho $. However, we consider the first three of these quantities fixed
by the protocol of interest and thus only display the dependence of the
success probability on the input density operator $\rho $. In the event of a
successful measurement, the output of the channels associated with $\mathcal{%
H}_{\bar{C}}$ is identified as the computational result, and it is described
by the reduced density operator 
\begin{equation}
\bar{\rho}=\frac{\mathrm{Tr}_{\bar{A}}\left( \bar{P}U\left( \rho \otimes
\sigma \right) U^{\dagger }\bar{P}\right) }{\mathrm{Tr}_{\bar{C},\bar{A}%
}\left( U\left( \rho \otimes \sigma \right) U^{\dagger }\bar{P}\right) }.
\label{rhobar}
\end{equation}
For any $\rho $ with $d(\rho )\neq 0$, this defines a so-called completely
positive (CP), trace preserving map $\mathcal{T}$ that takes each $\rho $ to
its associated $\bar{\rho}$: $\bar{\rho} = \mathcal{T}(\rho )$, relating
density operators in $\mathcal{H}_{C}$ to density operators in $\mathcal{H}_{%
\bar{C}}$. It will be convenient to write $\mathcal{T}(\rho )=\mathcal{V}%
(\rho )/d(\rho )$, where 
\begin{equation}
\mathcal{V}(\rho )\equiv \mathrm{Tr}_{\bar{A}}\left( \bar{P}U\left( \rho
\otimes \sigma \right) U^{\dagger }\bar{P}\right)  \label{Vdef}
\end{equation}
is a linear (non-trace preserving) CP map of density operators in $\mathcal{H%
}_{C}$ to positive operators in $\mathcal{H}_{\bar{C}}$ that is defined for
all density operators $\rho $ in $\mathcal{H}_{C}$. We restrict ourselves to
density operators $\rho $ over a subspace $\mathcal{S}_{C}$ of $\mathcal{H}%
_{C}$. This is usually the subspace in which the quantum gate operates.

As an example, consider the gate that turns the computational basis into the
Bell basis. In terms of polarization states, the subspace $\mathcal{S}_{C}$
might be spanned by the computational basis $\{|H,H\rangle ,|H,V\rangle
,|V,H\rangle ,|V,V\rangle \}$ (whereas $\mathcal{H}_{C}$ is spanned by the
full Fock basis). The Bell basis on $\mathcal{S}_{C}$ is then given by $\{
|\Psi^+\rangle, |\Psi^-\rangle, |\Phi^+\rangle, |\Phi^-\rangle \}$, where 
\[
|\Psi ^{\pm }\rangle =\frac{1}{\sqrt{2}}\left( |H,V\rangle \pm |V,H\rangle
\right) \text{~and~} |\Phi ^{\pm }\rangle =\frac{1}{\sqrt{2}}\left(
|H,H\rangle \pm |V,V\rangle \right) \;. 
\]
This gate is very important in quantum information theory, because it
produces maximal entanglement, and its inverse can be used to perform Bell
measurements. Both functions are necessary in, \textit{e.g.,} quantum
teleportation \cite{bennett93}. However, it is well known that such gates
cannot be constructed deterministically, and we therefore need to include an
ancilla state $\sigma$ and a projective measurement. We consider gates such
as these in this paper.

Suppose the subspace $\mathcal{S}_{C}$ is spanned by a set of vectors
labeled by Greek letters, \textit{e.g., }$\left| \alpha \right\rangle $. We
can then write 
\begin{equation}
\rho =\sum_{\alpha ,\beta }\left| \alpha \right\rangle \rho ^{\alpha \beta
}\left\langle \beta \right| ,  \label{rhodecompose}
\end{equation}
where $\rho ^{\alpha \beta }\equiv \left\langle \alpha |\rho |\beta
\right\rangle $. We identify a convex decomposition of the ancilla density
operator $\sigma $ as 
\[
\sigma =\sum_{i}p_{i}\left| \chi _{i}\right\rangle \left\langle \chi
_{i}\right| , 
\]
where the normalized (but not necessarily orthogonal) vectors $\left| \chi
_{i}\right\rangle $ are elements of $\mathcal{H}_{A}$, and the $p_{i}$ are
all non-negative and sum to unity, 
\[
\sum_{i}p_{i}=1\; . 
\]
We can then use (\ref{rhobar}) to write down an expression for the matrix
elements of $\bar{\rho}$. Note that it is possible to work with the
eigenkets of $\sigma $ so that $\left\{ \left| \chi _{i}\right\rangle
\right\} $ is an orthonormal set; however, this does not simplify the
analysis so we do not introduce the restriction. Furthermore, dealing with
non-orthogonal states in the ancilla convex decomposition may be more
convenient, depending on the system of interest. Choosing an orthonormal
basis of $\mathcal{H}_{\bar{C}}$ that we label by Greek letters with
overbars, \textit{e.g., }$\left| \bar{\alpha}\right\rangle $, we find 
\begin{equation}
\bar{\rho}^{\bar{\alpha}\bar{\delta}}=\sum_{\beta ,\gamma }\sum_{i,\bar{k}%
}\left( W_{\bar{k},i}^{\bar{\alpha}\beta }(\rho )\right) \rho ^{\beta \gamma
}\left( W_{\bar{k},i}^{\bar{\delta}\gamma }(\rho )\right) ^{*},
\label{rhobarwork}
\end{equation}
where 
\[
W_{\bar{k},i}^{\bar{\alpha}\beta }(\rho )=s_{\bar{k}}\sqrt{\frac{p_{i}}{%
d(\rho )}}\left( \left\langle \bar{k}\right| \left\langle \bar{\alpha}%
\right| \right) U\left( \left| \beta \right\rangle \left| \chi
_{i}\right\rangle \right) . 
\]
Note that 
\[
\sum_{\bar{\alpha}}\sum_{i,\bar{k}}\left( W_{\bar{k},i}^{\bar{\alpha}\gamma
}(\rho )\right) ^{*}\left( W_{\bar{k},i}^{\bar{\alpha}\beta }(\rho )\right)
=\delta _{\gamma \beta }\;, 
\]
which is confirmed by 
\begin{equation}
\mathrm{Tr}_{\bar{C}}(\bar{\rho})=\sum_{\bar{\alpha}}\bar{\rho}^{\bar{\alpha}%
\bar{\alpha}}=\sum_{\beta }\rho ^{\beta \beta }=\mathrm{Tr}_{C}(\rho ),
\label{tracepreserve}
\end{equation}
This last equation follows immediately from (\ref{rhobar}), since $\mathcal{T%
}$ is a trace preserving CP map and $\mathrm{Tr}_{C}(\rho )=1$.

In this paper, we consider a special class of maps that constitute a unitary
transformation on the computational subspace $\mathcal{S}_{C}$. In
particular, such transformations include the CNOT, the C-SIGN, and the
controlled bit flip. These are not the only useful maps in linear optical
quantum computing, but they arguably constitute the most important class.
Before we continue, we introduce the following definition:

\begin{description}
\item[Definition:]  We call a CP map $\rho \rightarrow \bar{\rho}=\mathcal{T}%
(\rho )$ an \textit{operationally unitary transformation }on density
operators $\rho $ over a subspace $\mathcal{S}_{C}$ if and only if:

\begin{itemize}
\item  For each $\rho $ over the subspace $\mathcal{S}_{C}$ we have $d(\rho
)\neq 0$, and

\item  For each $\rho $ defined by Eq.~(\ref{rhodecompose}) over the
subspace $\mathcal{S}_{C}$, the map $\mathcal{T}(\rho )$ yields a $\bar{\rho}
$ given by 
\begin{equation}
\bar{\rho}=\sum_{\alpha ,\beta }\left| \bar{\nu}_{\alpha }\right\rangle \rho
^{\alpha \beta }\left\langle \bar{\nu}_{\beta }\right| ,  \label{rhobareu}
\end{equation}
where the $\left| \bar{\nu}_{\alpha }\right\rangle $ are fixed vectors in $%
\mathcal{H}_{\bar{C}}$ satisfying $\left\langle \bar{\nu}_{\alpha }|\bar{\nu}%
_{\beta }\right\rangle =\left\langle \alpha |\beta \right\rangle =\delta
_{\alpha \beta }$.
\end{itemize}
\end{description}

This forms the obvious generalization of usual unitary evolution, since it
maintains the inner products of vectors under the transformation. Much of
our concern in this paper is in identifying the necessary and sufficient
conditions for a general map $\mathcal{T}(\rho )$ of Eqs.~(\ref{rhobar}) and
(\ref{rhobarwork}) to constitute an operationally unitary map. We begin in
the next section by considering what can be said about such maps.

\section{Consequences of operational unitarity}

\label{sec:consec}

In this section we restrict ourselves to CP maps $\mathcal{T}(\rho )$ that
are operationally unitary [see Eqs.~(\ref{rhodecompose}) and (\ref{rhobareu}%
)] for density operators $\rho $ over a subspace $\mathcal{S}_{C}$ of $%
\mathcal{H}_{C}$. The linearity of such maps implies that the convex sum of
two density operators is again a density operator: 
\[
\rho _{c}=x\rho _{a}+(1-x)\rho _{b}, 
\]
with $0\leq x\leq 1$. Applying Eqs.~(\ref{rhodecompose}) and (\ref{rhobareu}%
) to the three density operators $\rho _{a}$, $\rho _{b}$, and $\rho _{c}$
it follows immediately that 
\begin{equation}
\bar{\rho}_{c}=x\bar{\rho}_{a}+(1-x)\bar{\rho}_{b}.  \label{tranfirst}
\end{equation}
Now a second expression for $\bar{\rho}_{c}$ can be worked out by using the
defining relation (\ref{rhobar}) directly, 
\begin{eqnarray}
\bar{\rho}_{c} &=&\mathcal{T}(\rho _{c})=\frac{\mathcal{V}(\rho _{c})}{%
d(\rho _{c})}  \label{transecond} \\
&=&\frac{x\mathcal{V}(\rho _{a})+(1-x)\mathcal{V}(\rho _{b})}{xd(\rho
_{a})+(1-x)d(\rho _{b})}  \nonumber \\
&=&\frac{xd(\rho _{a})\bar{\rho}_{a}+(1-x)d(\rho _{b})\bar{\rho}_{b}}{%
xd(\rho _{a})+(1-x)d(\rho _{b})}  \nonumber
\end{eqnarray}
where in the second line we have used the linearity of $\mathcal{V}(\rho )$ (%
\ref{Vdef}) and $d(\rho )$ (\ref{Ddef}), and in the third line we have used
the corresponding relations for $\bar{\rho}_{a}$ in terms of $\rho _{a}$,
and $\bar{\rho}_{b}$ in terms of $\rho _{b}$. Setting the right-hand-sides
of Eqs.~(\ref{tranfirst}) and (\ref{transecond}) equal, we find 
\begin{equation}
x(1-x)\left[ d(\rho _{b})-d(\rho _{a})\right] (\bar{\rho}_{a}-\bar{\rho}%
_{b})=0.  \label{eureka}
\end{equation}
Since it is easy to see from Eqs.~(\ref{rhodecompose}) and (\ref{rhobareu})
that if $\rho _{a}$ and $\rho _{b}$ are distinct then $\bar{\rho}_{a}$ and $%
\bar{\rho}_{b}$ are as well; choosing $0<x<1$ it is clear that the only way
the operator equation (\ref{eureka}) can be satisfied is if $d(\rho
_{a})=d(\rho _{b})$. But since this must hold for \textit{any} two density
operators acting over $\mathcal{S}_{C}$, we have established that:

\begin{itemize}
\item  If a map $\mathcal{T}(\rho )$ is operationally unitary for $\rho $
(acting on a subspace $\mathcal{S}_{C}$), then $d(\rho )$ is independent of $%
\rho $: $d(\rho )=d$, for all $\rho $ acting on that subspace.
\end{itemize}

With this result in hand we can simplify Eq.~(\ref{rhobarwork}) for a map
that is operationally unitary, writing 
\begin{equation}
\bar{\rho}^{\bar{\alpha}\bar{\delta}}=\sum_{\beta ,\gamma }\sum_{J}w_{J}^{%
\bar{\alpha}\beta }\rho ^{\beta \gamma }\left( w_{J}^{\bar{\delta}\gamma
}\right) ^{*},  \label{tsimpform}
\end{equation}
where now 
\[
w_{J}^{\bar{\alpha}\beta }=w_{\bar{k},i}^{\bar{\alpha}\beta }=s_{\bar{k}}%
\sqrt{\frac{p_{i}}{d}}\left( \left\langle \bar{k}\right| \left\langle \bar{%
\alpha}\right| \right) U\left( \left| \beta \right\rangle \left| \chi
_{i}\right\rangle \right) 
\]
is independent of $\rho $; we have also introduced a single label $J$ to
refer to the pair of indices $\bar{k},i$. A further simplification arises
because the condition of operational unitarity guarantees that the subspace $%
\mathcal{S}_{\bar{C}}$ of $\mathcal{H}_{\bar{C}}$, over which the range of
density operators $\bar{\rho}$ generated by $\mathcal{T}(\rho )$ act as $%
\rho $ ranges over $\mathcal{S}_{C}$, has the same dimension as $\mathcal{S}%
_{C}$. We can thus adopt a set of orthonormal vectors $\left| \bar{\alpha}%
\right\rangle $ that span that subspace $\mathcal{S}_{\bar{C}}$, and the
matrices $w_{J}^{\bar{\alpha}\beta }$ are square.

At this point we can formally construct a unitary map on $\mathcal{S}_{C}$: $%
\tilde{\rho}\equiv \mathcal{U}(\rho )$, which is isomorphic in its effect on
density operators $\rho $ with our operationally unitary map $\bar{\rho}=%
\mathcal{T}(\rho )$. We do this by associating each $\left| \bar{\alpha}%
\right\rangle $ with the corresponding $\left| \alpha \right\rangle $,
introducing a density operator $\tilde{\rho}$ acting over $\mathcal{S}_{C}$,
and putting 
\begin{eqnarray}
\tilde{\rho}^{\alpha \delta } &\equiv &\bar{\rho}^{\bar{\alpha}\bar{\delta}},
\label{correspondence} \\
M_{J}^{\alpha \beta } &\equiv &w_{J}^{\bar{\alpha}\beta }.  \nonumber
\end{eqnarray}
The unitary map $\tilde{\rho}\equiv \mathcal{U}(\rho )$ is defined by the CP
map 
\[
\tilde{\rho}^{\alpha \delta }=\sum_{\beta ,\gamma }\sum_{J}M_{J}^{\alpha
\beta }\rho ^{\beta \gamma }(M_{J}^{\delta \gamma })^{*}, 
\]
or simply 
\begin{equation}
\tilde{\rho}=\sum_{J}M_{J}\rho M_{J}^{\dagger }.  \label{unitransform}
\end{equation}
This is often what is done implicitly when describing an operationally
unitary map, and we will see examples later in section \ref{sec:examples};
here we find this strategy useful to simplify our reasoning below.

Since the map $\tilde{\rho}\equiv \mathcal{U}(\rho )$ is unitary it can be
implemented by a unitary operator $M$, 
\[
\tilde{\rho}=M\rho M^{\dagger }, 
\]
where $M^{\dagger }=M^{-1}$. Thus $(M_{1},M_{2},....)$ and $(M,0,0,....)$,
where we add enough copies of the zero operator so that the two lists have
the same number of elements, constitute two sets of Kraus operators that
implement the same map $\tilde{\rho}\equiv \mathcal{U}(\rho )$. From Nielsen
and Chuang \cite{NandC} we have the following theorem:

\begin{description}
\item[Theorem:]  Suppose $\{E_{1},\ldots ,E_{n}\}$ and $\{F_{1},\ldots
,F_{m}\}$ are Kraus operators giving rise to CP linear maps $\mathcal{E}$
and $\mathcal{F}$ respectively. By appending zero operators to the shorter
list of elements we may ensure that $m=n$. Then $\mathcal{E}=\mathcal{F}$ if
and only if there exists complex numbers $u_{jk}$ such that $%
E_{j}=\sum_{k}u_{jk}F_{k}$, and $u_{jk}$ is an $m\times m$ unitary matrix.
\end{description}

Hence, $(M_{1},M_{2},....)$ must be related to $(M,0,0,....)$ by a unitary
matrix, and each $M_{J}$ is proportional to the single operator $M$. This
proof carries over immediately to the operationally unitary map $\mathcal{%
T(\rho )}$ under consideration, and we have

\begin{itemize}
\item  If a map $\mathcal{T}(\rho )$ is operationally unitary for $\rho $
acting over a subspace $\mathcal{S}_{C}$, then for fixed $\bar{k}$ and $i$
the square matrix defined by 
\[
w_{\bar{k},i}^{\bar{\alpha}\beta }=s_{\bar{k}}\sqrt{\frac{p_{i}}{d}}\left(
\left\langle \bar{k}\right| \left\langle \bar{\alpha}\right| \right) U\left(
\left| \beta \right\rangle \left| \chi _{i}\right\rangle \right) ,
\]
with $\bar{\alpha}$ labeling the row and $\beta $ the column, either
vanishes or is proportional to all other nonvanishing matrices identified by
different $\bar{k}$ and $i$. We can thus define a matrix $w^{\bar{\alpha}%
\beta }$ proportional to all the nonvanishing $w_{\bar{k},i}^{\bar{\alpha}%
\beta }$ such that we can write our map (\ref{tsimpform}) as 
\begin{equation}
\bar{\rho}^{\bar{\alpha}\bar{\delta}}=\sum_{\beta ,\gamma }w^{\bar{\alpha}%
\beta }\rho ^{\beta \gamma }\left( w^{\bar{\delta}\gamma }\right) ^{*}.
\label{tfinal}
\end{equation}
\end{itemize}

It is in fact easy to show that the two \textit{necessary} conditions we
have established here for a map $\mathcal{T}(\rho )$ to be an operationally
unitary transformation are also \textit{sufficient }conditions to guarantee
that it is. We show this in section \ref{sec:nscon}. First, however, we
establish a simple way of identifying whether or not $d(\rho )$ is
independent of $\rho $.

\section{The test condition}

In this section we consider a general map $\mathcal{T}(\rho )$ of the form
of Eq.~(\ref{rhobar}), and seek a simple condition equivalent to the
independence of $d(\rho )$ on $\rho $ for all $\rho $ acting over $\mathcal{S%
}_{C}$. To do this we write $d(\rho )$ of Eq.~(\ref{Ddef}) by taking the
complete trace over $\mathcal{H}_{C}$ and $\mathcal{H}_{A}$ rather than over 
$\mathcal{H}_{\bar{C}}$ and $\mathcal{H}_{\bar{A}}$, 
\begin{eqnarray*}
d(\rho ) &=&\mathrm{Tr}_{C,A}\left( U\left( \rho \otimes \sigma \right)
U^{\dagger }\bar{P}\right) \\
&=&\mathrm{Tr}_{C,A}\left( \left( \rho \otimes \sigma \right) U^{\dagger }%
\bar{P}U\right) \\
&=&\mathrm{Tr}_{C}(\rho\, T)
\end{eqnarray*}
where we have introduced a \textit{test operator }$T$ over the Hilbert space 
$\mathcal{H}_{C}$ as 
\[
T=\mathrm{Tr}_{A}\left( \sigma\, U^{\dagger }\bar{P}U\right) , 
\]
which does not depend on $\rho$. The operator $T$ is clearly Hermitian; it
is also a positive operator, since the probability for success $d(\rho )\geq
0$ for all $\rho $. We can now identify a condition for $d(\rho )$ to be
independent of $\rho$:

\begin{description}
\item[Theorem:]  $d(\rho )$ is independent of $\rho $, for density operators 
$\rho $ acting over a subspace $\mathcal{S}_{C}$ of $\mathcal{H}_{C}$, if
and only if the test operator $T\,$ is proportional to the identity operator 
${I}_{\mathcal{S}_{C}}$ over the subspace $\mathcal{S}_{C}.$ We refer to
this condition on $T$ as the \textit{test condition.}

\item[Proof:]  The sufficiency of the test condition for a $d(\rho )$
independent of $\rho $ is clear. Necessity is easily established by
contradiction: Suppose that $d(\rho )$ were independent of $\rho $ but $T$
not proportional to ${I}_{\mathcal{S}_{C}}$. Then at least two of the
eigenkets of $T$ must have different eigenvalues; call those eigenkets $%
\left| \mu _{a}\right\rangle $ and $\left| \mu _{b}\right\rangle $. It
follows that $d(\rho _{a})\neq d(\rho _{b})$, where $\rho _{a}=$ $\left| \mu
_{a}\right\rangle \left\langle \mu _{a}\right| $ and $\rho _{b}=\left| \mu
_{b}\right\rangle \left\langle \mu _{b}\right| $, in contradiction with our
assumption. \hfill $\square $
\end{description}

When the test condition is satisfied we denote the single eigenvalue of $T$
over $\mathcal{S}_{C}$ as $\tau $, i.e., $T=\tau {I}_{\mathcal{S}_{C}}$.
Then $d(\rho )=\tau $, and $\tau $ is identified as the probability that the
measurement indicated success. For any given protocol the calculation of the
operator $T$ gives an easy way to identify whether or not $d(\rho )$ is
independent of $\rho $.

\section{Necessary and sufficient conditions}

\label{sec:nscon}

We can now identify necessary and sufficient conditions for a map $\bar{\rho}%
=\mathcal{T}(\rho )$, to be an operationally unitary map for $\rho $ acting
on a subspace $\mathcal{S}_{C}$ of $\mathcal{H}_{C}$. They are:

\begin{enumerate}
\item  The test condition is satisfied: Namely, the operator 
\[
T=\mathrm{Tr}_{A}\left( \sigma \,U^{\dagger }\bar{P}U\right) 
\]
is proportional to the identity operator ${I}_{\mathcal{S}_{C}}$ over the
subspace $\mathcal{S}_{C}$.

\item  Each matrix 
\[
w_{\bar{k},i}^{\bar{\alpha}\beta }=s_{\bar{k}}\sqrt{\frac{p_{i}}{\tau }}%
\left( \left\langle \bar{k}\right| \left\langle \bar{\alpha}\right| \right)
U\left( \left| \beta \right\rangle \left| \chi _{i}\right\rangle \right) ,
\]
identified by the indices $\bar{k}$ and $i$, with row and column labels $%
\bar{\alpha}$ and $\beta $ respectively, either vanishes or is proportional
to all other such nonvanishing matrices; here $\tau $ is the eigenvalue of $T
$.
\end{enumerate}

The necessity of the first condition follows because it is equivalent to the
independence of $d(\rho )$ on $\rho $, which was established above as a
necessary condition for the transformation to be operationally unitary, as
was the second condition given here. So we need only demonstrate
sufficiency, which follows immediately: If the first condition is satisfied
then $d(\rho )=\tau $ is independent of $\rho $, and if the second is
satisfied then, from Eq.~(\ref{tsimpform}), we can introduce a single matrix 
$w^{\bar{\alpha}\beta }$ such that (\ref{tfinal}) is satisfied. Then 
\[
\sum_{\bar{\alpha}}\bar{\rho}^{\bar{\alpha}\bar{\alpha}}=\sum_{\beta ,\gamma
}\rho ^{\beta \gamma }\sum_{\bar{\alpha}}\left( w^{\bar{\alpha}\gamma
}\right) ^{*}w^{\bar{\alpha}\beta }. 
\]
Now the Hermitian matrix 
\[
Y^{\gamma \beta }\equiv \sum_{\bar{\alpha}}\left( w^{\bar{\alpha}\gamma
}\right) ^{*}w^{\bar{\alpha}\beta } 
\]
must in fact be the unit matrix: $Y^{\gamma \beta }=\delta _{\gamma \beta }$%
, otherwise we would not have 
\[
\sum_{\bar{\alpha}}\bar{\rho}^{\bar{\alpha}\bar{\alpha}}=\sum_{\beta }\rho
^{\beta \beta } 
\]
for an arbitrary $\rho $ over $\mathcal{S}_{C}$, and we know our general map 
$\bar{\rho}=\mathcal{T}(\rho )$ satisfies that condition [see Eq.~(\ref
{tracepreserve})]. Thus $w^{\bar{\alpha}\beta }$ is a unitary matrix, and
from the form of Eq.~(\ref{tfinal}) of the map from $\rho $ to $\bar{\rho}$
it follows immediately that the map is operationally unitary [see Eqs.~(\ref
{rhodecompose}) and (\ref{rhobareu})].

The physics of the two necessary and sufficient conditions given above is
intuitively clear, and indeed the results we have derived here could have
been guessed beforehand. For if the probability for success $d(\rho )$ of
the measurement were dependent of the input density operator $\rho $, by
monitoring the success rate in an assembly of experiments all characterized
by the same input $\rho $, one could learn something about $\rho $, and we
would not expect operationally unitary evolution in the presence of this
kind of gain of information. And the independence of the nonvanishing
matrices $w_{\bar{k},i}^{\bar{\alpha}\beta }$ on $\bar{k}$ and $i$, except
for overall factors, can be understood as preventing the `mixedness' of both
the input ancilla state $\sigma $ and the generally high rank projector $%
\bar{P}$, from degrading the operationally unitary transformation and
leading to a decrease in purity.

If a map is found to be operationally unitary, we can introduce the formally
equivalent unitary operator $M$ on $\mathcal{H}_{C}$, as in Eq.~(\ref
{correspondence}), which can then be written in terms of an effective action
operator $Q$, 
\begin{equation}
M=e^{-iQ/\hbar }\; .  \label{action}
\end{equation}
The operator $Q$ can be determined simply by diagonalizing $M$, and its form
reveals the nature of the Hamiltonian evolution simulated by the conditional
measurement process. We can define an effective Hamiltonian $H_{eff}$ that
characterizes an effective photon nonlinearity acting through a time $%
t_{eff}\,$by putting $H_{eff}\equiv Q/t_{eff}$, where $t_{eff}$ can be taken
as the time of operation of the device.

In a special but common case, the input ancilla state is pure and the
projector $\bar{P}$ is of unit rank in $\mathcal{H}_{\bar{A}}$. For cases
such as this there is only one matrix $w^{\bar{\alpha}\beta }$ in the
problem, and thus there is only a single necessary and sufficient condition
for the map to be operationally unitary:

\begin{itemize}
\item  In the special case of a projector $\bar{P}$ of rank 1 in $\mathcal{H}%
_{\bar{A}}$, where $\bar{P}=$ ${I}_{\bar{C}}\otimes \left| \bar{K}%
\right\rangle \left\langle \bar{K}\right| $, and a pure input ancilla state, 
$\sigma =\left| \chi \right\rangle \left\langle \chi \right| $, then map $%
\bar{\rho}=\mathcal{T}(\rho )$ is operationally unitary for $\rho $ acting
on a subspace $\mathcal{S}_{C}$ of $\mathcal{H}_{C}$ if and only if $T$
satisfies the test condition. Here 
\[
T=\left\langle \chi |U^{\dagger }\bar{P}U|\chi \right\rangle ,
\]
which is an operator in $\mathcal{H}_{C}$. If it does satisfy this
condition, then the transformation is given by 
\begin{equation}
\bar{\rho}^{\bar{\alpha}\bar{\delta}}=\sum_{\beta ,\gamma }w^{\bar{\alpha}%
\beta }\rho ^{\beta \gamma }\left( w^{\bar{\delta}\gamma }\right) ^{*},
\label{spectrans}
\end{equation}
where 
\[
w^{\bar{\alpha}\beta }=\sqrt{\frac{1}{\tau }}\left( \left\langle \bar{K}%
\right| \left\langle \bar{\alpha}\right| \right) U\left( \left| \beta
\right\rangle \left| \chi \right\rangle \right) ,
\]
and $\tau $ is the single eigenvalue of $T$ over $\mathcal{S}_{C}$.
\end{itemize}

\section{Generalization to include feed-forward processing}

\label{sec:post}

Suppose that the measurement outcome of the ancilla does not yield the
desired result, but that it signals that the output can be transformed by
simply applying a (deterministic) unitary mode transformation on the output
(see Fig. 2). This is called feed-forward processing
and is widely used. For example, in teleportation, Alice sends Bob a
classical message which allows him to correct for `wrong' outcomes of
Alice's Bell measurement. Here, we can explicitly take into account
feed-forward processing.

Suppose the projective measurement is characterized by a set of projectors,
each identifying a different detection signature, $\left\{ \bar{P}_{(1)},%
\bar{P}_{(2)},...\bar{P}_{(N)},\bar{P}_{\perp }\right\} $, where 
\[
\bar{P}_{\perp }={I-}\sum_{L=1}^{N}\bar{P}_{(L)},
\]
and 
\[
\bar{P}_{(L)}={I}_{\bar{C}}\otimes \sum_{\bar{k}}s_{L,\bar{k}}\left| \bar{k}%
\right\rangle \left\langle \bar{k}\right| .
\]
All the $s_{L,\bar{k}}$ are equal to zero or unity, such that 
\[
\bar{P}_{(L)}\bar{P}_{(L^{\prime })}=\bar{P}_{(L)}\delta _{LL^{\prime }}.
\]
Here success arises if the measurement outcome is associated with 
\textit{any }of the operators $\bar{P}_{(L)}$. And if outcome $L$ is
achieved, then the computational output is processed by application of
the unitary operator $\bar{V}_{(L)}$ acting over $\mathcal{H}_{\bar{C}}$.
The probability of achieving outcome $L$ is 
\[
d_{(L)}(\rho )\equiv \mathrm{Tr}_{\bar{C},\bar{A}}\left( U\left( \rho
\otimes \sigma \right) U^{\dagger }\bar{P}_{(L)}\right) 
\]
and if outcome $L$ is achieved the feed-forward processed computational output is
then 
\[
\bar{\rho}_{(L)}=\frac{\bar{V}_{(L)}\left[ \mathrm{Tr}_{\bar{A}}\left( \bar{P%
}_{(L)}U\left( \rho \otimes \sigma \right) U^{\dagger }\bar{P}_{(L)}\right)
\right] \bar{V}_{(L)}^{\dagger }}{\mathrm{Tr}_{\bar{C},\bar{A}}\left(
U\left( \rho \otimes \sigma \right) U^{\dagger }\bar{P}_{(L)}\right) }.
\]
which defines a map $\bar{\rho}_{(L)}=\mathcal{T}_{(L)}(\rho )$ for those $%
\rho $ for which $d_{(L)}(\rho )\neq 0$. In this more general case we define
the \textit{set} of maps $\left\{ \mathcal{T}_{(L)}\right\} $ to be
operationally unitary for density operators $\rho $ over the subspace $%
\mathcal{S}_{C}$ when:

\begin{itemize}
\item  For each $\rho $ over the subspace $\mathcal{S}_{C}$ at least one of
the $d_{(L)}(\rho )\neq 0$, and

\item  For each $\rho $ over the subspace $\mathcal{S}_{C}$, for each $L$
for which $d_{(L)}(\rho )\neq 0$ the map $\mathcal{T}_{(L)}(\rho )$ yields a 
$\bar{\rho}_{(L)}$ of the form of Eq.~(\ref{rhobareu}), independent of $L$.
\end{itemize}

The kind of arguments we have presented above can be extended to show that
the necessary and sufficient conditions for such a set of maps to be
operationally unitary for density operators $\rho $ over the subspace $%
\mathcal{S}_{C}$ are:

\begin{enumerate}
\item  Test conditions are satisfied: The operators 
\[
T_{(L)}=\mathrm{Tr}_{A}\left( \sigma U^{\dagger }\bar{P}_{(L)}U\right) 
\]
are each proportional to the identity operator ${I}_{\mathcal{S}C}$ over the
subspace $\mathcal{S}_{C}$. The proportionality constants $\tau _{(L)}$ need
not be the same for all $L$.

\item  Omitting matrices associated with any $L$ for which $\tau _{(L)}=0$,
each matrix 
\[
w_{L,\bar{k},i}^{\bar{\alpha}\beta }=s_{L,\bar{k}}\sqrt{\frac{p_{i}}{\tau
_{(L)}}}\sum_{\bar{\lambda}}\bar{V}_{(L)}^{\bar{\alpha}\bar{\lambda}}\left(
\left\langle \bar{k}\right| \left\langle \bar{\lambda}\right| \right)
U\left( \left| \beta \right\rangle \left| \chi _{i}\right\rangle \right) ,
\]
identified by the indices $L,\bar{k},$ and $i$, with row and column labels $%
\bar{\alpha}$ and $\beta $ respectively, either vanishes or is proportional
to all other such nonvanishing matrices.
\end{enumerate}

The probability of success is $\sum_{L}\tau _{(L)}=\tau .$ This expanded
formalism applies to the feed-forward schemes discussed by Pittman 
\textit{et al}.\ \cite{franson2} and the teleportation schemes of Gottesman
and Chuang \cite{gottesman}. In devices such as these, a measurement
provides classical information that is used in the subsequent evolution of
the output state.

In a common special case, the input ancilla state is pure, $\sigma =\left|
\chi \right\rangle \left\langle \chi \right| $, and each of the projectors $%
\bar{P}_{(L)}$ is of unit rank in $\mathcal{H}_{\bar{A}}$, $\bar{P}_{(L)}={I}%
_{\bar{C}}\otimes \left| \overline{k_{L}}\right\rangle \left\langle 
\overline{k_{L}}\right| $. Here the two necessary and sufficient conditions
for the set of maps to be operationally unitary for density operators $\rho $
over the subspace $\mathcal{S}_{C}$ simplify to:

\begin{enumerate}
\item  All the operators 
\[
T_{(L)}=\left\langle \chi |U^{\dagger }\bar{P}_{(L)}U|\chi \right\rangle 
\]
over $\mathcal{H}_{C}$ satisfy the test condition.

\item  Omitting matrices associated with any $L$ for which $\tau _{(L)}=0$,
each matrix 
\[
w_{L}^{\bar{\alpha}\beta }=\frac{1}{\sqrt{\tau _{(L)}}}\sum_{\bar{\lambda}}%
\bar{V}_{(L)}^{\bar{\alpha}\bar{\lambda}}\left( \left\langle \overline{k_{L}}%
\right| \left\langle \bar{\lambda}\right| \right) U\left( \left| \beta
\right\rangle \left| \chi \right\rangle \right) ,
\]
identified by the indices $L,$ with row and column labels $\bar{\alpha}$ and 
$\beta $ respectively, either vanishes or is proportional to all other such
nonvanishing matrices.
\end{enumerate}

If these conditions are met, then the operationally unitary transformation
is given by 
\[
\bar{\rho}^{\bar{\alpha}\bar{\delta}}=\sum_{\beta ,\gamma }w_{L}^{\bar{\alpha%
}\beta }\rho ^{\beta \gamma }\left( w_{L}^{\bar{\delta}\gamma }\right) ^{*},
\]
which is independent of $L$.

Another extension of the standard Von Neumann, or projection, measurements
is to the class of measurements described by more general positive operator
valued measures, or POVMs. \textit{\ }These can be used to describe more
complicated measurements, often resulting from imperfections in a designed
PVM. Our analysis can be generalized to POVMs by expanding the ancilla
space, and then describing the POVMs by PVMs in this expanded space. In some
instances operationally unitarity might still be possible; in others, the
extension would allow us to study of the effect of realistic limitations
such as detector loss and the lack of single-photon resolution.

\section{Examples}

\label{sec:examples}

In this section we will apply the formalism developed above to two proposed
optical quantum gates for LOQC. The straightforward calculation of the
effects of these gates presented in the original publications make it clear
that they are operationally unitary; our purpose here is merely to
illustrate how the approach we have introduced here is applied.

To evaluate the test operators $T_{(L)}$ and matrix elements $w_{L,\bar{k}%
,i}^{\bar{\alpha}\beta }$ it is useful to have expression for quantities
such as $Ua_{\Omega }U^{\dagger }$, where we use capital Greek letters as
subscripts on the letter $a$ to denote annihilation operators for input
(computing and ancilla) channels; similarly, we use $a_{\bar{\Delta}}$ to
denote annihilation operators for output (computing and ancilla) channels.
We now characterize the unitary transformation $U$ by a set of quantities 
\textsf{U} $_{\Omega \bar{\Delta}}^{*}$ that give the complex amplitude for
an output photon in mode $\bar{\Delta}$ given an input photon in mode $%
\Omega $. That is, 
\begin{equation}
U\left( a_{\Omega }^{\dagger }\left| \text{vac}\right\rangle \right) =\sum_{%
\bar{\Delta}}\mathsf{U}_{\Omega \bar{\Delta}}^{*}\left( a_{\bar{\Delta}%
}^{\dagger }\left| \text{vac}\right\rangle \right) ,  \label{Uadagger}
\end{equation}
where $\left| \text{vac}\right\rangle $ is the vacuum of the full Hilbert
space $\mathcal{H}$. Since only linear optical elements are involved we have 
$U^{\dagger }\left| \text{vac}\right\rangle =\left| \text{vac}\right\rangle $%
, and it further follows from (\ref{Uadagger}) that 
\begin{equation}
Ua_{\Omega }^{\dagger }U^{\dagger }=\sum_{\bar{\Delta}}\mathsf{U}_{\Omega 
\bar{\Delta}}^{*}a_{\bar{\Delta}}^{\dagger },  \label{Udaggerevol}
\end{equation}
or 
\begin{equation}
Ua_{\Omega }U^{\dagger }=\sum_{\bar{\Delta}}\mathsf{U}_{\Omega \bar{\Delta}%
}a_{\bar{\Delta}}.  \label{Ufund}
\end{equation}
Using the commutation relations satisfied by the creation and annihilation
operators, it immediately follows that the matrix $\mathsf{U}_{\Omega \bar{%
\Delta}}$ , which identifies the unitary transformation $U$, is itself a
unitary matrix. Certain calculations can be simplified by its
diagonalization, but for the kind of analysis of few photon states that we
require this is not necessary. We will need to express, in terms of few
photon states with respect to the decomposition $\mathcal{H}_{\bar{C}}$ $%
\otimes $ $\mathcal{H}_{\bar{A}}$, the result of acting with $U$ on few
photon states of the decomposition $\mathcal{H}_{C}$ $\otimes \mathcal{H}_{A}
$; this follows directly from (\ref{Udaggerevol}). For example, denoting by $%
\left| 1_{\Omega _{1}}2_{\Omega _{2}}\right\rangle $ the state with one
photon in mode $\Omega _{1}$ and two in mode $\Omega _{2}$, we have 
\begin{eqnarray}
U\left| 1_{\Omega _{1}}2_{\Omega _{2}}\right\rangle  &=&Ua_{\Omega
_{1}}^{\dagger }\frac{\left( a_{\Omega _{2}}^{\dagger }\right) ^{2}}{\sqrt{2}%
}\left| \text{vac}\right\rangle   \label{Uuse} \\
&=&\frac{1}{\sqrt{2}}\left( Ua_{\Omega _{1}}^{\dagger }U^{\dagger }\right)
\left( Ua_{\Omega _{2}}^{\dagger }U^{\dagger }\right) \left( Ua_{\Omega
_{2}}^{\dagger }U^{\dagger }\right) \left| \text{vac}\right\rangle  
\nonumber \\
&=&\frac{1}{\sqrt{2}}\sum_{\bar{\Delta}_{1},\bar{\Delta}_{2},\bar{\Delta}%
_{3}}\mathsf{U}_{\Omega _{1}\bar{\Delta}_{1}}^{*}\mathsf{U}_{\Omega _{2}\bar{%
\Delta}_{2}}^{*}\mathsf{U}_{\Omega _{2}\bar{\Delta}_{3}}^{*}\left( a_{\bar{%
\Delta}_{1}}^{\dagger }a_{\bar{\Delta}_{2}}^{\dagger }a_{\bar{\Delta}%
_{3}}^{\dagger }\right) \left| \text{vac}\right\rangle ,  \nonumber
\end{eqnarray}
and doing the sums in the last line allow us to indeed accomplish our goal.

\subsection{\protect\bigskip KLM conditional sign flip}

The first example we consider is the conditional sign flip discussed by
Knill, Laflamme, and Milburn \cite{klm}. Note that in this case the input
ancilla state is pure, there is no feed-forward processing, and the projector $\bar{P%
}$ is of unit rank in $\mathcal{H}_{\bar{A}}$. The necessary and sufficient
conditions for the transformation to be operationally unitary are those of
the special case discussed in section \ref{sec:nscon}. The gate consists of one
computational input port (labeled 1) and two ancilla input ports (2 and 3).
The projective measurement is performed on two output ports (b,c) and the
one remaining port is the computational output (a). The subspace $\mathcal{S}%
_{C}$ is spanned by the Fock states $|0\rangle$, $|1\rangle$, and $|2\rangle$
in each optical mode.

The pre-measurement evolution, which is done \textit{via }beam splitters and
a phase shifter, is given by the unitary transformation $U$ and
characterized by the matrix

\begin{equation}
\mathsf{U}=\mathsf{U}^{*}=\left[ 
\begin{array}{lll}
1-\sqrt{2} & 2^{-1/4} & (3/\sqrt{2}-2)^{1/2} \\ 
2^{-1/4} & 1/2 & 1/2-1/\sqrt{2} \\ 
(3/\sqrt{2}-2)^{1/2} & 1/2-1/\sqrt{2} & \sqrt{2}-1/2
\end{array}
\right] .
\end{equation}
The ancilla input state is 
\begin{equation}
\left| \chi \right\rangle =a_{2}^{\dagger }\left| \text{vac}%
_{A}\right\rangle ,  \label{ancilla input}
\end{equation}
denoting a single photon in the 2 mode, where $\left| \text{vac}%
_{A}\right\rangle $ denotes the vacuum of $\mathcal{H}_{A}$. The projective
measurement operator is given by

\[
\bar{P}={I}_{\bar{C}}\otimes \left| \bar{K}\right\rangle \left\langle \bar{K}%
\right| ={I}_{\bar{C}}\otimes a_{b}^{\dagger }\left| \text{vac}_{\bar{A}%
}\right\rangle \left\langle \text{vac}_{\bar{A}}\right| a_{b} 
\]
which corresponds to the detection of one and only one photon in mode b, and
zero photons in mode c. The basis states that define the subspace $\mathcal{S%
}_{C}$ are 
\[
\left| 0\right\rangle =\left| \text{vac}_{C}\right\rangle , \quad \left|
1\right\rangle =a_{1}^{\dagger }\left| \text{vac}_{C}\right\rangle , \quad
\left| 2\right\rangle =\frac{\left( a_{1}^{\dagger }\right) ^{2}}{\sqrt{2}}%
\left| \text{vac}_{C}\right\rangle , 
\]
and the basis states of $\mathcal{H}_{\bar{C}}$ are

\[
\left| \overline{0}\right\rangle =\left| \text{vac}_{\bar{C}}\right\rangle ,
\quad \left| \overline{1}\right\rangle =a_{a}^{\dagger }\left| \text{vac}_{%
\bar{C}}\right\rangle , \quad \left| \overline{2}\right\rangle =\frac{\left(
a_{a}^{\dagger }\right) ^{2}}{\sqrt{2}}\left| \text{vac}_{\bar{C}%
}\right\rangle , 
\]
In order to evaluate the test function, we first write 
\[
U^{\dagger }\bar{P}U=\sum_{\overline{\alpha }}U^{\dagger }\left(
a_{b}^{\dagger }\left| \text{vac}_{\bar{A}}\right\rangle \otimes \left| 
\overline{\alpha }\right\rangle \right) \left( \left\langle \overline{\alpha 
}\right| \otimes \left\langle \text{vac}_{\bar{A}}\right| a_{b}\right) U 
\]
and look at the matrix elements

\begin{eqnarray}
&&\left( \left\langle \alpha \right| \otimes \left\langle \chi\right|
\right) U^{\dagger }\bar{P}U\left( \left| \chi\right\rangle \otimes \left|
\beta \right\rangle \right)  \label{alphasum} \\
&=&\sum_{\overline{\alpha }}\left( \left\langle \alpha \right| \otimes
\left\langle \text{vac}_{A}\right| a_{2}\right) U^{\dagger }\left(
a_{b}^{\dagger }\left| \text{vac}_{\bar{A}}\right\rangle \otimes \left| 
\overline{\alpha }\right\rangle \right) \left( \left\langle \overline{\alpha 
}\right| \otimes \left\langle \text{vac}_{\bar{A}}\right| a_{b}\right)
U\left( a_{2}^{\dagger }\left| \text{vac}_{A}\right\rangle \otimes \left|
\beta \right\rangle \right)  \nonumber
\end{eqnarray}
over the computational subspace, $\mathcal{S}_{C}$. The calculation is
straightforward. Applying the operator $U$ on each of the states $%
a_{2}^{\dagger }\left| \text{vac}_{A}\right\rangle \otimes \left| \beta
\right\rangle $ gives the following states in the $\mathcal{H}_{\bar{C}}$ $%
\otimes $ $\mathcal{H}_{\bar{A}}$ decomposition

\begin{eqnarray*}
U\left( a_{2}^{\dagger }\left| \text{vac}_{A}\right\rangle \otimes \left|
0\right\rangle \right) &=&\left( 2^{-1/4}a_{a}^{\dagger }+\frac{1}{2}%
a_{b}^{\dagger }+\left[ \frac{1}{2}-\frac{1}{\sqrt{2}}\right] a_{c}^{\dagger
}\right) \left| \text{vac}\right\rangle \\
U\left( a_{2}^{\dagger }\left| \text{vac}_{A}\right\rangle \otimes \left|
1\right\rangle \right) &=&\left( 2^{-1/4}a_{a}^{\dagger }+\frac{1}{2}%
a_{b}^{\dagger }+\left[ \frac{1}{2}-\frac{1}{\sqrt{2}}\right] a_{c}^{\dagger
}\right) \\
&&\times \left( \left[ 1-\sqrt{2}\right] a_{a}^{\dagger
}+2^{-1/4}a_{b}^{\dagger }+\left[ \frac{3}{\sqrt{2}}-2\right]
^{1/2}a_{c}^{\dagger }\right) \left| \text{vac}\right\rangle \\
U\left( a_{2}^{\dagger }\left| \text{vac}_{A}\right\rangle \otimes \left|
2\right\rangle \right) &=&\frac{1}{\sqrt{2}}\left( 2^{-1/4}a_{a}^{\dagger }+%
\frac{1}{2}a_{b}^{\dagger }+\left[ \frac{1}{2}-\frac{1}{\sqrt{2}}\right]
a_{c}^{\dagger }\right) \\
&&\times \left( \left[ 1-\sqrt{2}\right] a_{a}^{\dagger
}+2^{-1/4}a_{b}^{\dagger }+\left[ \frac{3}{\sqrt{2}}-2\right]
^{1/2}a_{c}^{\dagger }\right) ^{2}\left| \text{vac}\right\rangle .
\end{eqnarray*}
and we can then separately evaluate the terms in the sum (\ref{alphasum}),
noting that the non-zero elements are 
\begin{eqnarray*}
\left| \left( \left\langle \overline{0}\right| \otimes \left\langle \text{vac%
}_{\bar{A}}\right| a_{b}\right) U\left( a_{2}^{\dagger }\left| \text{vac}%
_{A}\right\rangle \otimes \left| 0\right\rangle \right) \right| ^{2} &=&%
\frac{1}{4} \\
\left| \left( \left\langle \overline{1}\right| \otimes \left\langle \text{vac%
}_{\bar{A}}\right| a_{b}\right) U\left( a_{2}^{\dagger }\left| \text{vac}%
_{A}\right\rangle \otimes \left| 1\right\rangle \right) \right| ^{2} &=&%
\frac{1}{4} \\
\left| \left( \left\langle \overline{2}\right| \otimes \left\langle \text{vac%
}_{\bar{A}}\right| a_{b}\right) U\left( a_{2}^{\dagger }\left| \text{vac}%
_{A}\right\rangle \otimes \left| 2\right\rangle \right) \right| ^{2} &=&%
\frac{1}{4}
\end{eqnarray*}
The test operator $T$ is then

\begin{eqnarray*}
T &=&\frac{1}{4}\left[ \left| \text{vac}_{C}\right\rangle \left\langle \text{%
vac}_{C}\right| +a_{1}^{\dagger }\left| \text{vac}_{C}\right\rangle
\left\langle \text{vac}_{C}\right| a_{1}+\frac{\left( a_{1}^{\dagger
}\right) ^{2}}{\sqrt{2}}\left| \text{vac}_{C}\right\rangle \left\langle 
\text{vac}_{C}\right| \frac{\left( a_{1}\right) ^{2}}{\sqrt{2}}\right] \\
&=&\frac{1}{4}{I}_{\mathcal{S}C},
\end{eqnarray*}
and is indeed a multiple of the unit operator in the computational input
space. The probability of a success-indicating measurement is 1/4,
independent of the computational input state. Since this test condition is
satisfied, the transformation (\ref{spectrans}) is operationally unitary.
The terms of the transformation matrix $w^{\bar{\alpha}\beta }$ can be
calculated noting that the non-zero $\left\langle \bar{K}\right|
\left\langle \bar{\alpha}\right| U\left| \beta \right\rangle \left|
\chi\right\rangle $ terms are

\begin{eqnarray*}
\left\langle \bar{K}\right| \left\langle \overline{0}\right| U\left|
0\right\rangle \left| \chi\right\rangle &=&\frac{1}{2}, \\
\left\langle \bar{K}\right| \left\langle \overline{1}\right| U\left|
1\right\rangle \left| \chi\right\rangle &=&\frac{1}{2}, \\
\left\langle \bar{K}\right| \left\langle \overline{2}\right| U\left|
2\right\rangle \left| \chi\right\rangle &=&-\frac{1}{2},
\end{eqnarray*}
and since $\tau =1/4$ the non-zero elements of the transformation matrix are 
\begin{eqnarray*}
w^{\overline{0}0} &=&1, \\
w^{\overline{1}1} &=&1, \\
w^{\overline{2}2} &=&-1,
\end{eqnarray*}
which corresponds to the conditional sign flip, since with probability 1/4
the gate takes the input state $\left| \psi \right\rangle =\alpha _{0}\left|
0\right\rangle +\alpha _{1}\left| 1\right\rangle +\alpha _{2}\left|
2\right\rangle $ and produces the state $\left| \bar{\psi}\right\rangle
=\alpha _{0}\left| \overline{0}\right\rangle +\alpha _{1}\left| \overline{1}%
\right\rangle -\alpha _{2}\left| \overline{2}\right\rangle $.

This map can be seen to exhibit an effective nonlinear interaction between
the photons, since the formally equivalent unitary map (see section \ref{sec:consec}) is
characterized by the unitary operator $M$ (\ref{correspondence}), 
\[
\left| \tilde{\psi}\right\rangle =\alpha _{0}\left| 0\right\rangle +\alpha
_{1}\left| 1\right\rangle -\alpha _{2}\left| 2\right\rangle =M\left( \alpha
_{0}\left| 0\right\rangle +\alpha _{1}\left| 1\right\rangle +\alpha
_{2}\left| 2\right\rangle \right) ,
\]
which can be written in terms of an effective action operator $Q$ (\ref
{action}), where we can take 
\[
Q=\frac{\pi \hbar }{2}\left( 5\widehat{n}-\widehat{n}^{2}\right) ,
\]
with $\hat{n}$ the photon number operator. But such an effective action
operator exists only if we restrict ourselves to the three-dimensional
subspace $\mathcal{S}_{C}$, spanned by the kets $\left| 0\right\rangle $, $%
\left| 1\right\rangle $, and $\left| 2\right\rangle $. For consider an
attempt to expand this subspace to that spanned by the kets $\left( \left|
0\right\rangle ,\left| 1\right\rangle ,\left| 2\right\rangle ,\left|
3\right\rangle \right) $. The device guarantees that a computational input
of three photons can only produce a computational three-photon output, since
a successful measurement requires the detection of one and only one photon
in the ancilla space. The test operator is therefore still diagonal in the
photon number basis. However, we find 
\[
\left| \left( \left\langle \overline{3}\right| \otimes \left\langle \text{vac%
}_{\bar{A}}\right| a_{b}\right) U\left( a_{2}^{\dagger }\left| \text{vac}%
_{A}\right\rangle \otimes \left| 3\right\rangle \right) \right| ^{2}=\left( 2%
\sqrt{2}-\frac{5}{2}\right) ^{2},
\]
and thus the test operator $T$ is no longer a multiple of the unit operator
in this enlarged subspace. In this larger space the probability of a
success-indicating measurement is dependent on the input, and the map is not
operationally unitary.

\subsection{Polarization encoded CNOT}

The second example is the polarization-encoded Gottesman-Chuang protocol
discussed by Pittman \textit{et al}.\ \cite{franson}. In this case the input
ancilla state is pure, there is feed-forward processing, and there are several
projectors $\bar{P}_{(L)}$ of unit rank in $\mathcal{H}_{\bar{A}}$. The
necessary and sufficient conditions for the transformation to be
operationally unitary are therefore those of the special case discussed in
section \ref{sec:post}. The device has two computational input ports (labeled $a$ and $b$%
) and four ancilla input ports (1-4). A projective measurement is made on
four output ports ($p$,$q$,$n$,$m$) while the two remaining ports are the
computational output (5 and 6). A photon of horizontal polarization
represents a logical 0, and a vertically polarized photon represents a
logical 1. We use the same notation as Pittman \textit{et al. }\cite{franson}%
. For example, $\left| H(V)_{a}\right\rangle $ represents a
horizontally(vertically) polarized photon in port `a' and the Hadamard
transformed modes are $\left| F(S)_{a}\right\rangle =\frac{1}{2}\left[
\left| H_{a}\right\rangle \pm \left| V_{a}\right\rangle \right] $. The four
basis states of the computational input are $\left| 00\right\rangle =$ $%
\left| H_{a}\right\rangle \left| H_{b}\right\rangle ,$ $\left|
01\right\rangle =$ $\left| H_{a}\right\rangle \left| V_{b}\right\rangle ,$ $%
\left| 10\right\rangle =$ $\left| V_{a}\right\rangle \left|
H_{b}\right\rangle ,$ $\left| 11\right\rangle =$ $\left| V_{a}\right\rangle
\left| V_{b}\right\rangle $ and the output states are labeled as $\left| 
\overline{00}\right\rangle =\left| H_{5}\right\rangle \left|
H_{6}\right\rangle ,\left| \overline{01}\right\rangle =\left|
H_{5}\right\rangle \left| V_{6}\right\rangle ,\left| \overline{10}%
\right\rangle =\left| V_{5}\right\rangle \left| H_{6}\right\rangle ,\left| 
\overline{11}\right\rangle =\left| V_{5}\right\rangle \left|
V_{6}\right\rangle .$ The input ancilla state is 
\begin{eqnarray*}
\left| \chi \right\rangle  &=&\frac{1}{2}\left( \left| H_{1}\right\rangle
\left| H_{4}\right\rangle \left| H_{2}\right\rangle \left|
H_{3}\right\rangle +\left| H_{1}\right\rangle \left| V_{4}\right\rangle
\left| H_{2}\right\rangle \left| V_{3}\right\rangle \right)  \\
&&+\frac{1}{2}\left( \left| V_{1}\right\rangle \left| H_{4}\right\rangle
\left| V_{2}\right\rangle \left| V_{3}\right\rangle +\left|
V_{1}\right\rangle \left| V_{4}\right\rangle \left| V_{2}\right\rangle
\left| H_{3}\right\rangle \right) ,
\end{eqnarray*}
and the measurement projectors, $\bar{P}_{(L)}={I}_{\bar{C}}\otimes \left| 
\overline{k_{L}}\right\rangle \left\langle \overline{k_{L}}\right| ,$
represent the 16 possible success outcomes: 
\begin{eqnarray*}
\left| \overline{k_{1}}\right\rangle  &=&\left| F_{p}\right\rangle \left|
F_{q}\right\rangle \left| F_{n}\right\rangle \left| F_{m}\right\rangle  \\
&=&\frac{1}{4}\left( \left| H_{p}\right\rangle +\left| V_{p}\right\rangle
\right) \left( \left| H_{q}\right\rangle +\left| V_{q}\right\rangle \right)
\left( \left| H_{n}\right\rangle +\left| V_{n}\right\rangle \right) \left(
\left| H_{m}\right\rangle +\left| V_{m}\right\rangle \right)  \\
\left| \overline{k_{2}}\right\rangle  &=&\left| F_{p}\right\rangle \left|
F_{q}\right\rangle \left| F_{n}\right\rangle \left| S_{m}\right\rangle  \\
&=&\frac{1}{4}\left( \left| H_{p}\right\rangle +\left| V_{p}\right\rangle
\right) \left( \left| H_{q}\right\rangle +\left| V_{q}\right\rangle \right)
\left( \left| H_{n}\right\rangle +\left| V_{n}\right\rangle \right) \left(
\left| H_{m}\right\rangle -\left| V_{m}\right\rangle \right)  \\
&& \\
&&\vdots  \\
&& \\
\left| \overline{k_{15}}\right\rangle  &=&\left| S_{p}\right\rangle \left|
S_{q}\right\rangle \left| S_{n}\right\rangle \left| F_{m}\right\rangle  \\
&=&\frac{1}{4}\left( \left| H_{p}\right\rangle -\left| V_{p}\right\rangle
\right) \left( \left| H_{q}\right\rangle -\left| V_{q}\right\rangle \right)
\left( \left| H_{n}\right\rangle -\left| V_{n}\right\rangle \right) \left(
\left| H_{m}\right\rangle +\left| V_{m}\right\rangle \right)  \\
\left| \overline{k_{16}}\right\rangle  &=&\left| S_{p}\right\rangle \left|
S_{q}\right\rangle \left| S_{n}\right\rangle \left| S_{m}\right\rangle  \\
&=&\frac{1}{4}\left( \left| H_{p}\right\rangle -\left| V_{p}\right\rangle
\right) \left( \left| H_{q}\right\rangle -\left| V_{q}\right\rangle \right)
\left( \left| H_{n}\right\rangle -\left| V_{n}\right\rangle \right) \left(
\left| H_{m}\right\rangle -\left| V_{m}\right\rangle \right) 
\end{eqnarray*}
The polarizing beam splitters perform a unitary evolution on the input
ports, characterized by the set of quantities \textsf{U}$_{\Omega \bar{\Delta%
}}^{*}$. One can summarize the evolution of modes in $\mathcal{H}_{C}$ $%
\otimes$ $\mathcal{H}_{A}$ to modes in $\mathcal{H}_{\bar{C}}$ $\otimes $ $%
\mathcal{H}_{\bar{A}}$ with the following linear map 
\begin{eqnarray*}
\left| H_{1}\right\rangle  &\rightarrow &\left| H_{p}\right\rangle ,\left|
V_{1}\right\rangle \rightarrow -i\left| V_{q}\right\rangle , \\
\left| H_{2}\right\rangle  &\rightarrow &\left| H_{5}\right\rangle ,\left|
V_{2}\right\rangle \rightarrow \left| V_{5}\right\rangle , \\
\left| H_{3}\right\rangle  &\rightarrow &\left| H_{6}\right\rangle ,\left|
V_{3}\right\rangle \rightarrow \left| V_{6}\right\rangle , \\
\left| H_{4}\right\rangle  &\rightarrow &\left| H_{m}\right\rangle ,\left|
V_{4}\right\rangle \rightarrow -i\left| V_{n}\right\rangle , \\
\left| H_{a}\right\rangle  &\rightarrow &\left| H_{q}\right\rangle ,\left|
V_{a}\right\rangle \rightarrow -i\left| V_{p}\right\rangle , \\
\left| H_{b}\right\rangle  &\rightarrow &\left| H_{n}\right\rangle ,\left|
V_{b}\right\rangle \rightarrow -i\left| V_{m}\right\rangle ,
\end{eqnarray*}
since \textsf{U}$_{H_{1}H_{p}}^{*}=1,$ \textsf{U}$_{V_{1}V_{q}}^{*}=-i,$ 
\textit{etc}. As in the previous example, to evaluate the test operators, we
first look at the terms 
\begin{eqnarray*}
U\left( \left| 00\right\rangle \left| \chi \right\rangle \right)  &=&\frac{%
\left| H_{q}\right\rangle \left| H_{n}\right\rangle }{2}\left[ 
\begin{array}{c}
\left| H_{p}\right\rangle \left| H_{m}\right\rangle \left|
H_{5}\right\rangle \left| H_{6}\right\rangle -i\left| H_{p}\right\rangle
\left| V_{n}\right\rangle \left| H_{5}\right\rangle \left|
V_{6}\right\rangle  \\ 
-i\left| V_{q}\right\rangle \left| H_{m}\right\rangle \left|
V_{5}\right\rangle \left| V_{6}\right\rangle -\left| V_{q}\right\rangle
\left| V_{n}\right\rangle \left| V_{5}\right\rangle \left|
H_{6}\right\rangle 
\end{array}
\right]  \\
U\left( \left| 01\right\rangle \left| \chi \right\rangle \right)  &=&\frac{%
-i\left| H_{q}\right\rangle \left| V_{m}\right\rangle }{2}\left[ 
\begin{array}{c}
\left| H_{p}\right\rangle \left| H_{m}\right\rangle \left|
H_{5}\right\rangle \left| H_{6}\right\rangle -i\left| H_{p}\right\rangle
\left| V_{n}\right\rangle \left| H_{5}\right\rangle \left|
V_{6}\right\rangle  \\ 
-i\left| V_{q}\right\rangle \left| H_{m}\right\rangle \left|
V_{5}\right\rangle \left| V_{6}\right\rangle -\left| V_{q}\right\rangle
\left| V_{n}\right\rangle \left| V_{5}\right\rangle \left|
H_{6}\right\rangle 
\end{array}
\right]  \\
U\left( \left| 10\right\rangle \left| \chi \right\rangle \right)  &=&\frac{%
-i\left| V_{p}\right\rangle \left| H_{n}\right\rangle }{2}\left[ 
\begin{array}{c}
\left| H_{p}\right\rangle \left| H_{m}\right\rangle \left|
H_{5}\right\rangle \left| H_{6}\right\rangle -i\left| H_{p}\right\rangle
\left| V_{n}\right\rangle \left| H_{5}\right\rangle \left|
V_{6}\right\rangle  \\ 
-i\left| V_{q}\right\rangle \left| H_{m}\right\rangle \left|
V_{5}\right\rangle \left| V_{6}\right\rangle -\left| V_{q}\right\rangle
\left| V_{n}\right\rangle \left| V_{5}\right\rangle \left|
H_{6}\right\rangle 
\end{array}
\right]  \\
U\left( \left| 11\right\rangle \left| \chi \right\rangle \right)  &=&\frac{%
-\left| V_{p}\right\rangle \left| V_{m}\right\rangle }{2}\left[ 
\begin{array}{c}
\left| H_{p}\right\rangle \left| H_{m}\right\rangle \left|
H_{5}\right\rangle \left| H_{6}\right\rangle -i\left| H_{p}\right\rangle
\left| V_{n}\right\rangle \left| H_{5}\right\rangle \left|
V_{6}\right\rangle  \\ 
-i\left| V_{q}\right\rangle \left| H_{m}\right\rangle \left|
V_{5}\right\rangle \left| V_{6}\right\rangle -\left| V_{q}\right\rangle
\left| V_{n}\right\rangle \left| V_{5}\right\rangle \left|
H_{6}\right\rangle 
\end{array}
\right] 
\end{eqnarray*}
The matrix elements of interest are now

\begin{eqnarray}
&&\left( \left\langle \alpha \right| \otimes \left\langle \chi \right|
\right) U^{\dagger }\bar{P}_{(L)}U\left( \left| \chi \right\rangle \otimes
\left| \beta \right\rangle \right)   \label{alphasum2} \\
&=&\sum_{\overline{\alpha }}\left( \left\langle \alpha \right| \otimes
\left\langle \chi \right| \right) U^{\dagger }\left| \overline{k_{L}}%
\right\rangle \left| \overline{\alpha }\right\rangle \left\langle \overline{%
\alpha }\right| \left\langle \overline{k_{L}}\right| U\left( \left| \chi
\right\rangle \otimes \left| \beta \right\rangle \right)   \nonumber
\end{eqnarray}
and the non-zero terms of the sum in (\ref{alphasum2}) are 
\begin{eqnarray*}
\left| \left\langle \overline{00}\right| \left\langle \overline{k_{L}}%
\right| U\left( \left| \chi \right\rangle \otimes \left| 00\right\rangle
\right) \right| ^{2} &=&\frac{1}{16} \\
\left| \left\langle \overline{01}\right| \left\langle \overline{k_{L}}%
\right| U\left( \left| \chi \right\rangle \otimes \left| \overline{01}%
\right\rangle \right) \right| ^{2} &=&\frac{1}{16} \\
\left| \left\langle \overline{11}\right| \left\langle \overline{k_{L}}%
\right| U\left( \left| \chi \right\rangle \otimes \left| \overline{10}%
\right\rangle \right) \right| ^{2} &=&\frac{1}{16} \\
\left| \left\langle \overline{10}\right| \left\langle \overline{k_{L}}%
\right| U\left( \left| \chi \right\rangle \otimes \left| \overline{11}%
\right\rangle \right) \right| ^{2} &=&\frac{1}{16}
\end{eqnarray*}
for all $L$. The test functions, $\left\{ T_{(L)}\right\} $ , are then

\begin{eqnarray*}
T_{(L)} &=&\frac{1}{64}\left[ 
\begin{array}{c}
\left| H_{a}\right\rangle \left| H_{b}\right\rangle \left\langle
H_{b}\right| \left\langle H_{a}\right| +\left| H_{a}\right\rangle \left|
V_{b}\right\rangle \left\langle V_{b}\right| \left\langle H_{a}\right|  \\ 
+\left| V_{a}\right\rangle \left| H_{b}\right\rangle \left\langle
H_{b}\right| \left\langle V_{a}\right| +\left| V_{a}\right\rangle \left|
V_{b}\right\rangle \left\langle V_{b}\right| \left\langle V_{a}\right| 
\end{array}
\right]  \\
&=&\frac{1}{64}\;{I}_{\mathcal{S}C}
\end{eqnarray*}
and are indeed multiples of the unit operator in the computational input
space. In this scheme $\tau _{(L)}=1/64$, and the probability of success is
the sum of the individual probabilities of the 16 detection outcomes, $%
\sum_{L}\tau _{(L)}=1/4.$ The terms of the transformation matrices $w_{L}^{%
\bar{\alpha}\beta }$ can be calculated noting that the non-zero $%
\left\langle \overline{k_{L}}\right| \left\langle \bar{\lambda}\right|
U\left| \beta \right\rangle \left| \chi \right\rangle $ terms are 
\begin{eqnarray*}
\left\langle \overline{k_{L}}\right| \left\langle \overline{00}\right|
U\left| 00\right\rangle \left| \chi \right\rangle  &=&e^{i\phi _{L,0}}/8, \\
\left\langle \overline{k_{L}}\right| \left\langle \overline{01}\right|
U\left| 01\right\rangle \left| \chi \right\rangle  &=&e^{i\phi _{L,1}}/8, \\
\left\langle \overline{k_{L}}\right| \left\langle \overline{11}\right|
U\left| 10\right\rangle \left| \chi \right\rangle  &=&e^{i\phi _{L,2}}/8, \\
\left\langle \overline{k_{L}}\right| \left\langle \overline{10}\right|
U\left| 11\right\rangle \left| \chi \right\rangle  &=&e^{i\phi _{L,3}}/8,
\end{eqnarray*}
where $e^{i\phi _{L,0}}=1,e^{i\phi _{1,1}}=-1,e^{i\phi _{2,1}}=1,\ldots
,e^{i\phi _{16,3}}=1$ are phase factors of $\pm $1$.$ For this
transformation to be operationally unitary, the $w_{L}^{\bar{\alpha}\beta }$
matrices must all be proportional to each other. In certain outcomes,
single-qubit operations ($\pi $-phase shifts) are required to correct the
phase factors so that the transformation is operationally unitary and the
desired output is produced. The feed-forward processing matrices, $\bar{V}_{(L)}^{%
\bar{\alpha}\bar{\lambda}}$, represent these single qubit operations.
Setting 
\[
\bar{V}_{(L)}^{\overline{00},\overline{00}}=e^{i\phi _{L,0}},\quad \bar{V}%
_{(L)}^{\overline{01},\overline{01}}=e^{i\phi _{L,1}},\quad \bar{V}_{(L)}^{%
\overline{11},\overline{11}}=e^{i\phi _{L,2}},\quad \bar{V}_{(L)}^{\overline{%
10},\overline{10}}=e^{i\phi _{L,3}}\;,
\]
with all other elements equal to zero gives the appropriate corrections. The
non-zero transformation matrix elements, are then 
\begin{eqnarray*}
w_{L}^{\overline{00},00} &=&1 \\
w_{L}^{\overline{01},01} &=&1 \\
w_{L}^{\overline{11},10} &=&1 \\
w_{L}^{\overline{10},11} &=&1
\end{eqnarray*}
for all $L$. Since the 16 evolution matrices are identical, the
proportionality condition is satisfied. The transformation is then 
\[
\bar{\rho}^{\bar{\alpha}\bar{\delta}}=\sum_{\beta ,\gamma }w_{L}^{\bar{\alpha%
}\beta }\rho ^{\beta \gamma }\left( w_{L}^{\bar{\delta}\gamma }\right) ^{*}
\]
which is the CNOT operation. This gate takes the input state $\left| \psi
\right\rangle =\alpha _{0}\left| 00\right\rangle +\alpha _{1}\left|
01\right\rangle +\alpha _{2}\left| 10\right\rangle +\alpha _{3}\left|
11\right\rangle $ and produces the state $\alpha _{0}\left| \overline{00}%
\right\rangle +\alpha _{1}\left| \overline{01}\right\rangle +\alpha
_{2}\left| \overline{11}\right\rangle +\alpha _{3}\left| \overline{10}%
\right\rangle $ with probability 1/4. Again, this map exhibits an effective
nonlinear interaction between the photons since the formally equivalent
unitary map is characterized by a nonlinear effective action operator $Q$ (%
\ref{action}). In this case one could choose 
\[
Q=\frac{\pi \hbar }{2}(3+a_{b}^{\dagger }(1-\hat{n}_{b})+(1-\hat{n}%
_{b})a_{b})\hat{n}_{a}.
\]
Again, however, the operational unitarity is restricted to the subspace.
Suppose we expand the computational subspace to include an extra photon in
one of the input modes. As an example, consider the special state $\left|
S\right\rangle =$ $\left| H_{a}\right\rangle \left| H_{b}\right\rangle
\left| H_{b}\right\rangle $. The form of the projectors indicates that the
detection events involve one and only one photon in the appropriate modes.
Evaluating the corresponding test operator elements we find 
\[
\left| \left\langle \overline{\alpha }\right| \left\langle \overline{k_{L}}%
\right| U\left( \left| \chi \right\rangle \otimes \left| S\right\rangle
\right) \right| ^{2}=0\;,
\]
since the extra photon inhibits a success-indicating measurement result. The
evolution cannot be operationally unitary in this expanded subspace because
the test operator is no longer proportional to the unit operator.

\section{Conclusion}

\label{sec:conclusions}

In this paper we introduced a general approach to the investigation of
conditional measurement devices. We considered an important class of optical 
$N$-port devices, including those employing projectors of rank greater than
unity, mixed input ancilla states, multiple success outcomes, and
feed-forward processing. We also sketched how more general POVMs, rather than PVMs,
could be included. The necessary and sufficient conditions for these devices
to simulate unitary evolution have been derived. They are not surprising,
and indeed from a physical point of view are fairly obvious. But to our
knowledge they have not been discussed in this general way before. One of
the conditions is that the probability of each successful outcome must be
independent of the input density operator. Whether or not this holds can be
checked by evaluating a set of test operators over the input computational
Hilbert space, which is easily done for any proposed device. In the special
case of only one successful outcome there is only one test operator to be
computed; furthermore, if the ancilla state is pure and the success
projector of rank one, then the passing of a test condition by that single
test operator guarantees that the map is operationally unitary. In the case
of more than one successful outcome it is a necessary consequence of
operational unitarity that each of the test operators pass the test
condition. This is not sufficient to imply operational unitarity in the
multiple projector case unless the proportionality condition is also
satisfied. The proportionality condition can often be satisfied by
introducing feed-forward processing.

Besides application in the analysis of particular proposed devices, we
believe the general framework presented here will be useful in exploring the
different types of pre-measurement evolution and measurements that might be
useful in the design, optimization, and characterization of such devices. In
particular, the conditional sign flip and polarization-encoded CNOT devices
we considered functioned as operationally unitary maps only over the input
computational subspaces for which they were originally proposed. So while
effective photon nonlinearities could be introduced, the degree to which
they are physically meaningful is somewhat limited. An outstanding issue,
perhaps even of interest more from the general perspective of nonlinear
optics than from that of quantum computer design, is the study of potential
devices that provide effective photon nonlinearities over much larger input
computational subspaces. The question remains: to what extent are such
devices possible in theory and feasible in practice?

Finally, we note that only in section \ref{sec:examples} did we assume that
the pre-measurement unitary evolution $U$ is associated with linear elements
in an optical system. The more general framework of the earlier sections may
find application in describing other proposed devices for quantum
information processing that involve conditional measurement schemes in the
presence of more complicated interactions \cite{dfs}.

\section*{Acknowledgements}

This work was supported by the Natural Sciences and Engineering Research
Council of Canada and the Walter C. Sumner Foundation. Part of this work was
carried out at the Jet Propulsion Laboratory, California Institute of
Technology, under a contract with the National Aeronautics and Space
Administration. In addition, P.K.\ acknowledges the United States National
Research Council. Support was also received from the Advanced Research and
Development Activity, the National Security Agency, the Defense Advanced
Research Projects Agency, and the Office of Naval Research. We would
like to thank Alexei Gilchrist, James Franson, and Gerard Milburn for
stimulating discussions.

\pagebreak
\begin{picture}(400,150)(0,0)
\put(150,0){\framebox(60,120){U}}
\put(75,90){\vector(1,0){75}}
\put(75,30){\vector(1,0){75}}
\put(210,90){\vector(1,0){75}}
\put(210,30){\vector(1,0){25}}
\put(235,10){\framebox(70,40)}
\put(30,75){Computational Input}
\put(220,75){Computational Output}
\put(40,15){Ancilla Input}
\put(240,33){Measurement}
\put(260,18){on $\mathcal{H}_{\bar{A}}$}
\put(290,90){$\mathcal{H}_{\bar{C}}$}
\put(55,30){$\mathcal{H}_{A}$}
\put(55,90){$\mathcal{H}_{C}$}
\end{picture}
\begin{figure}[h!]
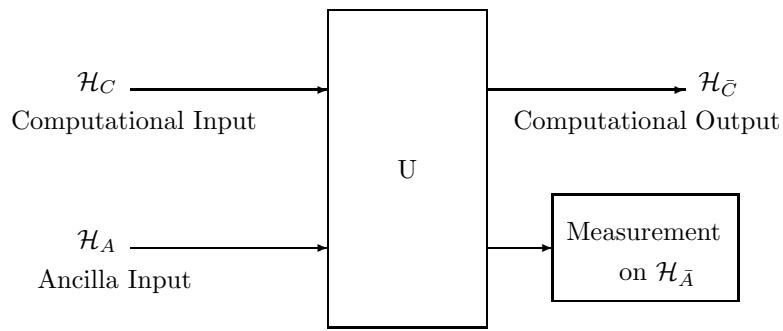

\caption{A schematic diagram of a basic conditional measurement
device.
The input computational channels, $\mathcal{H}_{C}$, and input ancilla
channels, $\mathcal{H}_{A}$,
undergo unitary evolution.
The measurement performed in the ancilla output space,
$\mathcal{H}_{\bar{A}}$, indicates the success
or failure of the computation.}
\end{figure}

\begin{picture}(400,200)
\put(120,0){\framebox(60,120){U}}
\put(75,90){\vector(1,0){45}}
\put(75,30){\vector(1,0){45}}
\put(275,90){\vector(1,0){20}}
\put(180,30){\vector(1,0){25}}
\put(180,90){\vector(1,0){25}}
\put(205,70){\framebox(70,40)}
\put(205,10){\framebox(70,40)}
\put(210,33){Measurement}
\put(229,18){on $\mathcal{H}_{\bar{A}}$}
\put(212,93){Feed-forward}
\put(212,78){processing}
\put(300,90){$\mathcal{H}_{\bar{C}}$}
\put(55,30){$\mathcal{H}_{A}$}
\put(55,90){$\mathcal{H}_{C}$}
\put(242,50){\line(0,1){20}}
\put(238,50){\line(0,1){20}}
\end{picture}
\begin{figure}[h!]
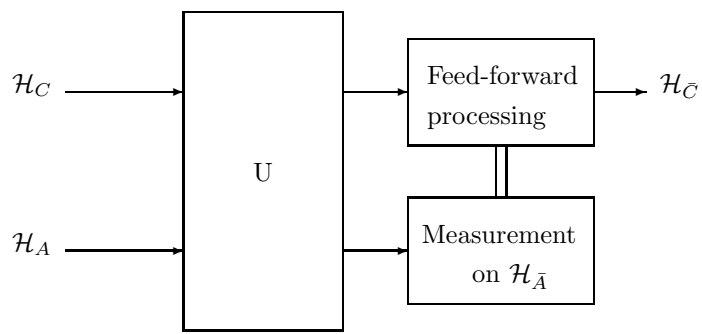

\caption{A schematic diagram of a conditional measurement
device that incorporates
feed-forward processing.  The double line connecting the two small boxes
represents a classical channel that carries the measurement result.  Based
on the outcome, the appropriate processing
is performed on the output channel.}
\end{figure}

\end{document}